\documentclass[12pt]{iopart}

\usepackage{graphicx}

\usepackage{cite}

\newcommand{\xx}{\mathbf{x}}

\newcommand{\pp}{\mathbf{p}}
\newcommand{\rr}{\mathbf{r}}
\newcommand{\nn}{\mathbf{n}}
\newcommand{\zz}{\mathbf{a}}
\newcommand{\ZZ}{\mathbf{b}}

\newcommand{\beq}{\begin{eqnarray}}
\newcommand{\eeq}{\end{eqnarray}}

\begin{document}

\title[Epidemic outbreak distributions]{WKB calculation of an epidemic outbreak distribution}

\author{Andrew J Black$^1$ and Alan J McKane$^2$}

\address{$^1$ School of Mathematical Sciences, The University of Adelaide, 
Adelaide, SA 5005, Australia}

\address{$^2$ Theoretical Physics Division, School of Physics and Astronomy, 
University of Manchester, Manchester M13 9PL, UK}

\ead{andrew.black@adelaide.edu.au}

\begin{abstract}
We calculate both the exponential and pre-factor contributions in a WKB 
approximation of the master equation for a stochastic SIR model with highly 
oscillatory dynamics. Fixing the basic parameters of the model we investigate 
how the outbreak distribution changes with the population size. We show that 
this distribution rapidly becomes highly non-Gaussian, acquiring large tails 
indicating the presence of rare, but large outbreaks, as the population is 
made smaller. The analytic results are found to be in excellent agreement with 
simulations until the systems become so small that the dynamics are dominated 
by fade-out of the disease.

\end{abstract}


\submitto{J. Stat. Mech.}

\maketitle

\section{Introduction}

The dynamics of recurrent epidemics are an active topic of research in a wide 
range of disciplines: from ecology and epidemiology to applied mathematics and 
theoretical physics~\cite{Bau08,MBG+09b,Rozhnova2010}. Much of the reason for 
this can be attributed to the large amount of historical data 
available~\cite{bartlett2,REG99}, and the rich variety of highly oscillatory 
patterns that are observed in both the data and relatively simple stochastic 
models \cite{bauch_earn:inter,KR07}.

The systematic formulation of individually-based models of epidemics starting 
from a master equation (continuous-time Markov chain) has permitted a parallel
analytical and numerical study to be carried out, which has allowed a rather
complete understanding of these stochastic models to be achieved. One of the 
main features of these models is that demographic stochasticity tends to 
excite the natural dynamics of the system leading to large scale coherent 
oscillations, termed stochastic amplification~\cite{alonso}. It has been shown 
how the frequency and amplitude of these oscillations depend on the parameters 
of the system and can provide a parsimonious explanation for the range of 
frequencies observed in real data~\cite{BE03,Black2010b,Black2010,Rozhnova2010}. In all of these studies though, the population is assumed to be very large 
so low numbers of infectives, or fade-out of the disease, is unlikely.

\begin{figure}[ht]
\centering
\includegraphics[width=0.7\textwidth]{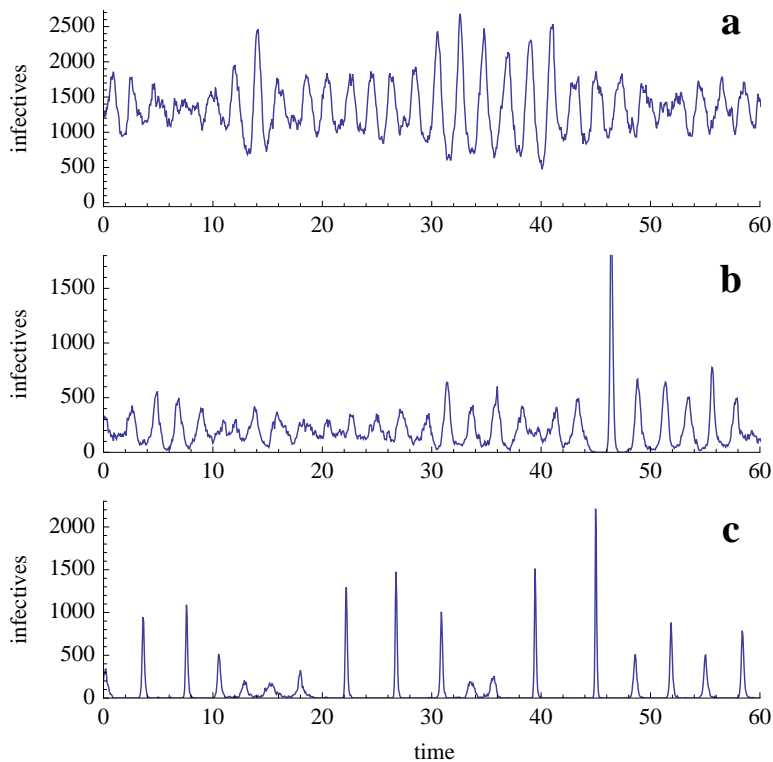}
\caption{Infective time series illustrating how the dynamics of the model 
change as the total population, $N$, is decreased. (a) $N=2\times 10^6$, the 
number of infectives oscillates about the mean, but fade-out does not happen. 
(b) $N=3 \times 10^5$, fade-out does occur but rarely. (c) $N=10^5$, here 
outbreaks are mainly triggered by immigration events and rapidly lead to 
fade-out. The model and parameters which generate these are described in 
Sections 2 and 3.} 
\label{fig:time_series}
\end{figure}

For childhood diseases especially, where the infectious period is orders of 
magnitude smaller than the lifetime~\cite{and_may}, the size of the epidemic 
outbreaks can be so large that the number of infectives can become small and 
fade-out of the disease is very probable, even in medium-size 
populations~\cite{bartlett2}. Because of this, some form of migration is 
commonly included in these models to re-introduce the disease after fade-out. 
This is a biologically realistic feature which is not needed in deterministic 
models where fade-out cannot happen. The presence of this fade-out boundary in 
the stochastic model means there is a natural asymmetry in the state space, 
which can have a considerable effect on the dynamics of the fluctuations in 
smaller-to-medium size systems. Figure \ref{fig:time_series} illustrates the 
dynamics of a stochastic epidemic model at three different population sizes. 
As the population is decreased the probability of a large outbreak, relative 
to the mean infection level, increases. These smaller systems are important 
as they are representative of the majority of towns and cities for which 
extensive case-report data exists~\cite{Grenfell1998}, but have so far 
received little theoretical attention.

In this paper we calculate analytically---via a Wentzel-Kramers-Brillouin (WKB)
approximation of the master equation---the outbreak distribution (marginal 
infective probability distribution) of a susceptible-infected-recovered (SIR) 
model. We show how the asymmetry in the system affects this distribution and 
in particular how it changes with population size. In the limit $N\to\infty$, 
the distribution tends to a Gaussian, as is expected and used in the 
system-size expansion of the master equation~\cite{van_kampen}. As the 
population is made smaller the distribution becomes skewed, rapidly acquiring 
fat-tails indicating an increased probability of rare, but large, outbreaks. 
All results are compared with stochastic simulations and, within the 
approximation's range of validity, the agreement is excellent.  Although we
have used the term analytic to refer to the method which we use, it should be 
understood by this that we use analytic techniques to obtain equations or 
expressions for functions making up probability distributions, but  
numerical techniques have to be used to solve these to obtain the final 
results.

The WKB approximation has a long history in the theory of stochastic 
processes, and was the basis of Kramer's calculation of escape of a particle 
over a potential barrier due to thermal noise~\cite{Kramers1940}. The method 
was originally used on Fokker-Planck equations with continuous variables
$\mathbf{x}$ and consisted of making an ansatz for the probability of the 
form $P(\mathbf{x}) \sim \exp{[-NS(\mathbf{x})]}$, where $N$ was a large 
parameter such a the volume of the system or the inverse diffusion constant.
The function $S(\mathbf{x})$ turns out to obey a Hamilton-Jacobi 
equation~\cite{Goldstein}, and and so leads to a definition of a Hamiltonian 
dynamics which describes the stochastic dynamics of rare events, such as large 
outbreaks.

For the situation discussed in this paper we require a variant of the method 
applied to the master equation, where the variables are discrete rather than 
continuous~\cite{Kubo1973,Gang1987,Dykman1994}. However the method is similar 
and the problem of describing strongly stochastic effects near the 
fade-out boundary is transformed into one of classical mechanics and of finding 
zero energy trajectories of a Hamiltonian. It should be stressed that this 
deterministic dynamics is not simply the $N \to \infty$ dynamics of the 
original stochastic model, but an auxiliary dynamics which is able to give 
a very good description of processes that cannot be described by the 
conventional deterministic dynamics found by taking $N \to \infty$.

The method allows higher-order corrections to the leading 
$\exp{[-NS(\mathbf{x})]}$ to be calculated, the next-order correction being a
prefactor $K(\mathbf{x})$ multiplying the 
exponential~\cite{Stein1997,Roma2005}. There have been a  number of 
applications of this method in the last few years, for example to calculate 
fixation and extinction times from quasi-stationary to absorbing 
states~\cite{Schwartz2009,MS09,Kamenev2008,Khasin2010}, or transition times 
between metastable states~\cite{Dykman1994,Escudero2009,Assaf2010}. The WKB 
method is especially suitable for these type of problems as the aim is to 
only find one special trajectory. The work in this paper differs from these 
previous studies in that we are primarily concerned in studying the dynamics 
of the model, and hence in calculating the full stationary distribution of the 
master equation, not fixation times (there is no fixation in our model since 
immigration is included). Another distinguishing aspect of our work is that 
we also calculate the pre-factor term for the multi-dimensional model, which 
as will be shown, is crucial for accurate calculation of the probability 
density. This has not so far been computed for a model of this type.

The rest of this paper is organised as follows: in Section \ref{sec:WKB} we 
introduce the theory behind the WKB approximation of the master equation and in
Section \ref{sec:solution} describe the numerical procedure that we adopt to 
apply it to calculate approximate distributions. In Section \ref{sec:results} 
we apply these methods to the SIR model with immigration. We end with a 
discussion of our results and outline a number of open questions.


\section{General formalism}
\label{sec:WKB}
In this section we will discuss the WKB approximation to the master equation 
and its application to the SIR model. Although much of the formalism we 
require appears in various places in the literature it is somewhat scattered, 
so we first give a coherent summary based on some of the clearer discussions 
that have been published \cite{Dykman1994,Stein1997,Roma2005}.

The starting point is the master equation for the time-evolution of
$P_\mathbf{n}(t)$, which is the probability of finding the system in state
$\mathbf{n}$ at time $t$:
\beq
\frac{dP_{\mathbf{n}}(t)}{dt} = \sum_{\mathbf{r}} 
\left[ T_{\mathbf{r}}(\mathbf{n}-\mathbf{r})P_{\mathbf{n}-\mathbf{r}}(t)
- T_{\mathbf{r}}(\mathbf{n})P_{\mathbf{n}}(t) \right]\,.
\label{full_master}
\eeq
Here $\mathbf{r}$ labels the transitions by giving the size of the jump. Both
$\mathbf{n}$ and $\mathbf{r}$ are vectors of integers; for the SIR model they
are two-dimensional with $\mathbf{n}=(n,m)$, where $n$ and $m$ are the number 
of susceptible and infected individuals respectively. Setting the left-hand 
side of \Eref{full_master} to zero we find the time-independent equation 
specifying the stationary distribution,
\beq
\sum_{\mathbf{r}} \left[ T_{\mathbf{r}}(\mathbf{n}-\mathbf{r})
P_{\mathbf{n}-\mathbf{r}} - T_{\mathbf{r}}(\mathbf{n})P_{\mathbf{n}}\right]
= 0\,.
\label{master}
\eeq
We now expand the transition rates in terms of the system size, $N\gg1$. In our
case this is the number of individuals in the system. Specifically we write
\beq
T_{\mathbf{r}}(\mathbf{n}) =  N W_{\mathbf{r}}(\mathbf{n}) +  
U_{\mathbf{r}}(\mathbf{n}) + {\cal O}(1/N)\,,
\label{expand_rates}
\eeq
where $W_{\mathbf{r}}(\mathbf{n})$ and $U_{\mathbf{r}}(\mathbf{n})$ are of order
unity. For the SIR Model, the transition rates are such that 
$U_{\mathbf{r}}(\mathbf{n})$ and the higher-order corrections are zero, and 
so $T_{\mathbf{r}} =  N W_{\mathbf{r}}$.

Finally we introduce the fraction of individuals who fall into the different 
classes: $\mathbf{x} = \mathbf{n}/N$; for the SIR model we will write 
$\mathbf{x}=(x,y)$ where $x=n/N$ and $y=m/N$. So letting
$W_{\mathbf{r}}(\mathbf{n})=W_{\mathbf{r}}(N\mathbf{x})=w_{\mathbf{r}}(\mathbf{x})$
and also $P_{\mathbf{n}}\equiv P_{N\xx}=\pi(\mathbf{x})$, the master equation 
(\ref{master}) now reads
\beq
\sum_{\mathbf{r}} \left[ w_{\mathbf{r}}(\mathbf{x}-\frac{\mathbf{r}}{N})
\pi(\mathbf{x}-\frac{\mathbf{r}}{N}) - w_{\mathbf{r}}(\mathbf{x})
\pi(\mathbf{x})\right]=0\,.
\label{final_master}
\eeq


\subsection{The WKB approximation}
\label{sec:gen_form}
We can now apply the WKB approximation to $\pi(\mathbf{x})$ by expressing it as
\beq
\pi(\mathbf{x})=K(\mathbf{x})\exp{\left( -NS(\mathbf{x}) \right)}\left[ 1 + 
{\cal O}\left(\frac{1}{N}\right) \right]\,,
\label{WKB_ansatz}
\eeq
where both $S(\mathbf{x})$ and $K(\mathbf{x})$ are assumed to be of order 
unity and $N\gg1$. Next we expand $S(\mathbf{x}-(\mathbf{r}/N))$ to second 
order:
\beq\fl
S(\xx-\frac{\rr}{N}) = S(\xx)-\frac{1}{N}\rr\cdot\nabla_\xx S(\xx) + 
\frac{1}{2N^2}\left( {\rr}\cdot{\nabla_\xx} \right)^{2}S(\xx) +
{\cal O}\left(\frac{1}{N^3}\right),
\label{secondorder_S}
\eeq
and $K(\xx)$ to first order: 
$K(\xx-{\rr}/N) = K(\xx)- \rr\cdot\nabla_\xx K(\xx)/N + {\cal O}(1/N^2)$. 
Substituting into the master equation and comparing terms, at leading order 
we find a Hamilton-Jacobi equation, corresponding to the Hamiltonian
\beq
H(\mathbf{x},\mathbf{p})=\sum_\mathbf{r} w_\rr(\mathbf{x})
[\exp(\mathbf{r}\cdot\mathbf{p})-1]=0.
\label{eq:ham}
\eeq
Here $\pp=\nabla_\xx S(\xx)$. The prefactor $K(\mathbf{x})$ is determined by 
the next-order contributions in $N^{-1}$ and is discussed in Section 
\ref{prefactor}. It should not be too surprising that a Hamilton-Jacobi 
equation is found using the form (\ref{WKB_ansatz}): Markovian stochastic 
processes are equivalent to quantum mechanics in imaginary time~\cite{Risken}
and we are interested in the limit where $N^{-1}$ ($\hbar$) goes to zero. 

Hamilton's equations are found from Eq.~(\ref{eq:ham}) to be:
\beq
\dot{\mathbf{x}} &=& \nabla_\mathbf{p} H=\sum_\rr \rr\,w_\rr(\xx) 
\exp{({\rr}\cdot{\pp})}\,, \nonumber \\
\dot{\pp} &=& -\nabla_\xx H = -\sum_\rr[\exp{({\rr}\cdot{\pp})}-1]
\nabla_\xx w_\rr(\xx)\,,
\label{eq:motion}
\eeq
where the dot denotes differentiation with respect to time. From the solution
of these equations we will find trajectories, called the fluctuational 
trajectories, $\xx_f(t)$ and the ``momenta'' along these trajectories 
$\pp_f(t)$. For the zero-energy solutions that we are interested in, the 
action along a given fluctuational trajectory is given by
\beq
S_f = \int_{t_0}^t {\pp}_{f}\cdot{\dot{\xx}_{f}}\,dt'\,.
\label{eq:action}
\eeq

\noindent
From Eq.~(\ref{eq:motion}) we see that $\pp_{f}=0$ is always one solution,
with
\beq
\dot{\mathbf{x}} = \sum_\rr \rr\,w_\rr(\xx)\,.
\label{relax}
\eeq
This is called the relaxation trajectory for the following reason. If we
multiply the original master equation (\ref{full_master}) by $\nn$, sum over
all $\nn$, and then shift the first sum by $\rr$, we find that 
$d\langle \nn \rangle/dt=\sum_{\rr} \rr\,T_{\rr}(\langle{\nn}\rangle)$, in the 
limit $N \to \infty$. Introducing $\xx$ and $w_{\rr}(\xx)$ as above, gives
exactly Eq.~(\ref{relax}). Therefore Eq.~(\ref{relax}) describes the actual 
deterministic dynamics of the system which would occur in the limit 
$N \to \infty$. This is not the trajectory of primary interest to us in
this paper; instead our focus will be on solutions of Eq.~(\ref{eq:motion}) 
with $\pp_{f}\neq0$, which correspond to non-allowed trajectories in the
$N \to \infty$ limit. Before we discuss these, we will apply this formalism to
the SIR model.


\subsection{The SIR model with immigration}
\label{sec:SIR}
The SIR model consists of three classes of individuals, susceptibles denoted
by $S$, infected denoted by $I$ and recovered denoted by $R$ \cite{and_may}. 
We assume that births and deaths are coupled at the individual level, so that 
when an individual dies another (susceptible) individual is born \cite{alonso}.
This means that the number of individuals, $N$, does not change with time, 
and so the number of recovered individuals is not an independent variable:
$n_{\rm R}=N-n-m$. Immigration is included with a commuter formalism where 
susceptibles are assumed to be in contact with a constant pool of infectives 
outside the main community \cite{alonso,KR07}. 

The transitions in the model are then of the following type:
\begin{enumerate}
\item Infection: $S+I \stackrel {\beta}{\longrightarrow } I+I$ and 
$S \stackrel {\eta}{\longrightarrow} I$.
\[ T(n-1,m+1|n,m) =\Bigl(\beta\frac{n}{N}m + \eta n\Bigr).
\]
\item Recovery: $I \stackrel{\gamma}{\longrightarrow} R$.
\[
T(n,m-1|n,m)=\gamma m. 
\] 
\item Death of an infected individual: $I \stackrel{\mu}{\longrightarrow} S$.
\[
T(n+1,m-1|n,m)=\mu m. 
\]
\item Death of a recovered individual: $R \stackrel {\mu}{\longrightarrow} S$.
\[
T(n+1,m|n,m)=\mu(N-n-m). 
\]
\end{enumerate}

Expressing this model in terms of the quantities introduced at the start of 
this section, we find that, as already mentioned, all corrections to 
$NW_{\rr}(\nn)$ in Eq.~(\ref{expand_rates}) are zero, and that $w_{\rr}(\xx)$ 
is given by
\begin{itemize}
\item[(i)] Infection: $w_\rr(\xx)=\beta x y +\eta x, \quad\quad \rr=(-1,+1)$
\vspace{0.2cm}
\item[(ii)] Recovery: $w_\rr(\xx)=\gamma y, \quad\quad \rr=(0,-1)$
\vspace{0.2cm}
\item[(iii)] Death of an infected individual: 
$w_\rr(\xx)=\mu y, \quad\quad \rr=(+1,-1)$
\vspace{0.2cm}
\item[(iv)] Death of a recovered individual: 
$w_\rr(\xx)=\mu (1-x-y), \quad\quad \rr=(+1,0)$
\end{itemize}
Substituting these rates into (\ref{eq:ham}) we find the Hamiltonian 
\beq
H(\xx,\pp)&=&(\beta x y+\eta x)[e^{-p_x+p_y}-1]+\gamma y [e^{-p_y}-1]\nonumber \\
&+& \mu y [e^{-p_y+p_x}-1]+\mu(1-x-y)[e^{p_x}-1]\,.
\label{Ham_SIR}
\eeq

\noindent
Related Hamiltonians have been used by previous authors 
\cite{Kamenev2008,Schwartz2009}. The relaxation trajectory can be found by 
setting $\pp = 0$ in the Hamilton equations found from \Eref{Ham_SIR}, or by 
simply substituting the $w_\rr(\xx)$ for the SIR model into \Eref{relax}. 
Either way this gives the deterministic SIR equations:
\beq
\dot{x}&=&-\beta x y -\eta x+\mu(1-x),\nonumber \\
\dot{y}&=&\beta x y +\eta x -(\gamma+\mu)y\,.
\label{SIR_deter}
\eeq
From \Eref{SIR_deter} we see that the fixed points (denoted by an 
asterisk) satisfy $y^*=\mu(1-x^*)/(\gamma+\mu)$. Eliminating $y^*$ shows that 
$x^*$ may be found by solving
\beq
(x^*)^2-\left[1+\frac{\eta}{\mu}\frac{(\gamma+\mu)}{\beta}+
\frac{(\gamma+\mu)}{\beta}\right]x^*+\frac{(\gamma+\mu)}{\beta}=0\,.
\label{x_star}
\eeq
This equation has two fixed points: a `trivial' one which vanishes if $\eta=0$ 
and a non-trivial (endemic) one. This is the one which will be of interest to 
us in the rest of the paper. The focus will be on the fluctuational 
trajectories which after a perturbation to this fixed point at $t=0$, rapidly 
move away and end near the extinction boundary~\cite{Dykman1994,MS09}.


\subsection{Calculation of the Prefactor $K(\xx)$}
\label{prefactor}
An equation for the prefactor $K(\xx)$ is found from examining the second-order
${\cal O}(N^{-1})$ terms in the expansion of the master equation. From this 
we find the equation
\beq
\sum_\rr w_\rr\,\exp({\rr}\cdot{\pp}) \left[ \frac{{\rr}\cdot{\nabla_\xx K}}{K}
+\frac{1}{2} ({\rr}\cdot{\nabla_\xx})^2 S + 
\frac{{\rr}\cdot{\nabla_\xx w_\rr}}{w_\rr} \right] = 0.
\label{first_form}
\eeq
If $U_r\neq0$ in the expansion of the transition rates \eref{expand_rates}, 
then there are extra terms in this equation~\cite{Assaf2010}. Using Hamilton's 
equations~(\ref{eq:motion}), and their derivatives, we may write 
\Eref{first_form} in the alternative form
\beq
\frac{1}{K}\sum_i \frac{\partial H}{\partial p_i}
\frac{\partial K}{\partial x_i} + 
\frac{1}{2}\sum_{i,j} \frac{\partial^2 S}{\partial x_i \partial x_j} 
\frac{\partial^2 H}{\partial p_i \partial p_j} + 
\sum_i \frac{\partial^2 H}{\partial p_i \partial x_i} = 0\,.
\label{second_form}
\eeq
Since $K$ has no explicit time dependence, we have that
\beq
\frac{dK}{dt} = \sum_i \dot{x}_i \frac{\partial K}{\partial x_i} =
\sum_i \frac{\partial H}{\partial p_i}\frac{\partial K}{\partial x_i}\,.
\label{total_deri}
\eeq
Using this result in Eq.~(\ref{second_form}) we find a differential equation 
for $K$,
\beq
\frac{dK}{dt} = -\left[ \sum_i \frac{\partial^2 H}{\partial p_i \partial x_i}+
\frac{1}{2}\sum_{i,j} \frac{\partial^2 S}{\partial x_i \partial x_j} 
\frac{\partial^2 H}{\partial p_i \partial p_j}\right]K.
\label{eq:amp_evo}
\eeq
To find $K(\xx)$ from this equation, we need to know the Hessian, 
$Z_{ij} \equiv \partial^2 S/\partial x_i \partial x_j$. It is important to 
realise that we are only interested in quantities which vary along the 
trajectories in phase space found by solving Hamilton's equations, so the
momenta are now functions of $\xx$. Differentiating $H=0$ twice with respect to 
$\xx$ taking care to distinguish between the cases where $\xx$ and $\pp$ are 
independent and when they are not, one finds the following equation for the
matrix $Z$~\cite{Maier1993,Stein1997,Roma2005,Bressloff2010}:
\beq\fl
\dot{Z}_{ij}+\sum_{k,l}\frac{\partial^2H}{\partial p_k \partial p_l}Z_{ik}Z_{jl}+
\sum_{l} \frac{\partial^2H}{\partial x_i \partial p_l}Z_{jl}+
\sum_{l} \frac{\partial^2H}{\partial x_j \partial p_l}Z_{il}+
 \frac{\partial^2H}{\partial x_i \partial x_j}=0\,.
\label{eq:hessian_evo}
\eeq
This can be written as a matrix equation
\beq
\dot{Z}+Z H_{\pp\pp}Z+H_{\xx\pp}Z +Z H_{\pp\xx}+H_{\xx\xx}=0\,,
\label{matrix_eqn}
\eeq
where $H_{\zz\ZZ}=\partial_\zz \partial_\ZZ H$, with $\zz$ and $\ZZ$ being $\xx$
or $\pp$.


\subsection{Gaussian asymptote}
\label{normalisation}
To actually solve Hamilton's equations and thus calculate $S(\xx)$ and $K(\xx)$
we need to know boundary conditions. The fact that the stationary distribution 
is approximately Gaussian near the fixed point, $\xx^*$, can provide 
these~\cite{van_kampen,Dykman1994}. From \Eref{eq:amp_evo} we see that the 
differential equation which determines $K$ is linear, and so although the 
dependence on $\xx$ can be found from this equation, $K$ will only be known 
up to an overall constant. 

From Eq.~(\ref{eq:action}) we see that $S(\mathbf{x}^*)=0$ and so the WKB 
approximation (\ref{WKB_ansatz}) can equivalently be written as 
\beq
\pi(\mathbf{x})=K(\mathbf{x})\exp{\left( -N \left[S(\mathbf{x})
- S(\mathbf{x}^*) \right] \right)}\left[ 1 + {\cal O}(\frac{1}{N}) \right]\,.
\label{WKB_ansatz_mod}
\eeq
Expanding about the fixed point for large $N$ to Gaussian order gives
\beq
\pi(\mathbf{x})=K(\mathbf{x}^*)\exp{\left[-\frac{N}{2}
(\xx-\xx^*)^TZ^{*}(\xx-\xx^*)\right]}\left[ 1 + {\cal O}(\frac{1}{N}) \right]\,.
\label{Gaussian_approx}
\eeq
The value of $K(\xx^*)$ now follows by asking that the probability distribution
is normalised. Carrying out the Gaussian integrals one finds that
\beq 
K(\xx^*) = \frac{2N}{\pi}\,{\rm det}Z^{*}\,.
\label{K_star}
\eeq
The matrix $Z$ at the fixed point, 
$\xx = \xx^*$ and $\pp = 0$, can be found from Eq.~(\ref{matrix_eqn}). Since 
$\dot{Z}=0$ and 
\beq
& & 
\left.\frac{\partial^2 H}{\partial x_i \partial p_j}\right|_{\pp =0, \, \xx=\xx^*}
= \sum_\rr r_j \left. \frac{\partial w_\rr(\xx)}{\partial x_i}\right|_{\xx=\xx^*}
\equiv A_{ji} \nonumber \\
& & 
\left.\frac{\partial^2 H}{\partial p_i \partial p_j}\right|_{\pp =0, \, \xx=\xx^*}
= \sum_\rr w_\rr(\xx^*)r_i r_j \equiv B_{ij} \nonumber \\
& & 
\left.\frac{\partial^2 H}{\partial x_i \partial x_j}\right|_{\pp =0, \, \xx=\xx^*}
= 0\,,
\label{Z_FP}
\eeq
we have from Eq.~(\ref{matrix_eqn}) that $Z^{*}BZ^{*}+A^{\rm T}Z^{*}+Z^{*}A=0$,
or 
\beq
B+\Xi A^{\rm T}+A \Xi=0,
\label{B_X_Xi}
\eeq
where $\Xi$ is the covariance matrix and $Z^{*}\equiv \Xi^{-1}$. It should be 
noted that the Gaussian result \eref{Gaussian_approx} is what would be 
obtained from the system-size expansion to first order \cite{van_kampen}.


\section{Solution procedure}
\label{sec:solution}
In this section we describe how we calculate the stationary probability 
density from the WKB approximation. As we are primarily concerned with how 
the form of the stationary distribution changes with $N$, we fix the basic 
parameters as: $\beta=1.3$, $\gamma=1/13$, $\mu=5.5\times 10^{-5}$ and 
$\eta=10^{-6}$. These correspond to the childhood disease measles and lead to 
highly oscillatory stochastic dynamics~\cite{and_may} (see Figure 
\ref{fig:time_series}). The effect of changing these parameters is discussed 
later in Section~\ref{sec:dis}.

\begin{figure}[ht]
\centering
\includegraphics[width=\textwidth]{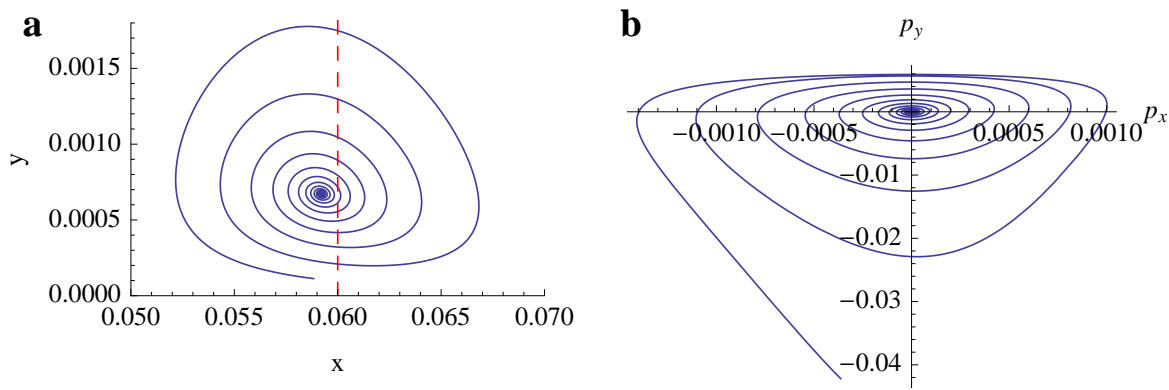}
\caption{An example of a fluctuational trajectory emanating from the fixed 
point $\xx=\xx^*$, $\pp=0$, projected into the (a) $x$-$y$ plane, and (b) 
$p_x$-$p_y$ plane. The line $x=0.06$ is marked by the red dashed line.}
\label{fig:trajectory}
\end{figure}

The procedure starts with the Hamiltonian for the SIR model (\ref{Ham_SIR}). 
The problem is to find the fluctuational solutions of Hamilton's equations 
(\ref{eq:motion}). As already discussed, there is a single stable fixed point 
at $\xx=\xx^*$, $\pp=0$, and the non-trivial fluctuational trajectories that 
we seek have $\pp\neq 0$. We can use the results of Section \ref{normalisation}
to derive a consistent set of initial conditions to equations \eref{eq:motion} 
by noticing that $S(\xx)$ is quadratic in $\xx-\xx^*$ near the fixed 
point~\cite{Dykman1994}. If we define a small perturbation, $\delta x_j$, to 
the fixed point, then the corresponding initial perturbations to the momentum 
along the required trajectory are given by
\beq
\delta p_i = \sum_j 
\left.\frac{\partial p_i}{\partial x_j}\right|_{\pp =0,\,\xx=\xx^*} \delta x_j =
\sum_j Z^{*}_{ij} \delta x_j =\sum_j (\Xi^{-1})_{ij} \delta x_{j}.
\label{eq:mtm_pert}
\eeq
The covariance matrix $\Xi$ may be found from \Eref{B_X_Xi}, since $A$ and $B$ 
can be found from \Eref{Z_FP}. An example of a fluctuational trajectory found  
via this procedure is shown in Figure \ref{fig:trajectory}. The trajectory 
spirals away from the fixed point and ends when it leaves the positive quadrant
of the $x$-$y$ plane. Some care has to be taken when specifying the initial 
perturbations; they must be small enough that \Eref{eq:mtm_pert} applies, but 
not so small that numerical errors can build up along the trajectory, due to 
the exponential slow down near the fixed point. For the same reason there is 
a practical limit to the range over which trajectories, and hence $\pi(\xx)$ 
can be calculated with this method.

With the initial conditions, plus $S(\xx^*)=0$, $K(\xx^*)=1$ and 
$Z(\xx^*)=\Xi^{-1}$, we can simply numerically integrate forward the equations 
\eref{eq:motion}, \eref{eq:action}, \eref{eq:amp_evo} and \eref{eq:hessian_evo}
simultaneously to find a trajectory as well as $S(\xx)$ and $K(\xx)$ along it. 
We have taken $K(\xx^*)=1$, rather than the value given by Equation 
(\ref{K_star}) for convenience. The final results are always normalised 
numerically, so the choice we make here is not important. By changing the 
initial perturbation, $\delta x_j$, different trajectories can be found. As 
the functions $S$ and $K$, and hence $\pi$, are only given along the 
trajectory, we need a method to interpolate these into a distribution over 
the $x$-$y$ plane. 

\begin{figure}[ht]
\centering
\includegraphics[width=\textwidth]{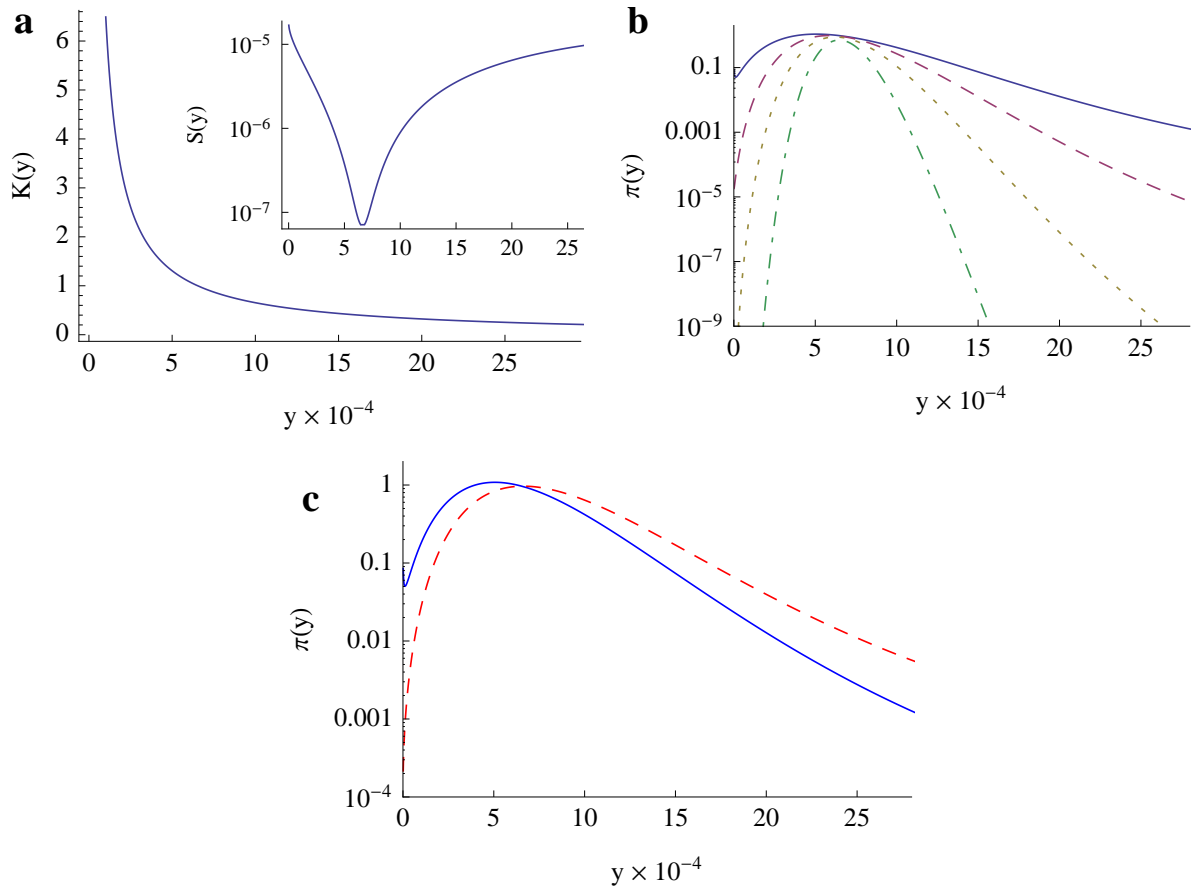}
\caption{(a) The functions $K$ and $S$ evaluated along the line $x=0.06$ (see 
Figure \ref{fig:trajectory}a). (b) The probability density, $\pi(y)=K(y)\exp[-NS(y)]$, 
evaluated for $N=5\times10^6$, $2\times10^6$, $10^6$ and $5\times10^5$ (solid, 
dashed, dotted and dot-dashed lines respectively). (c) Comparison of $\pi(y)$ 
calculated with (solid blue curve) and without (red dashed curve) the 
pre-factor $K(y)$; $N=5\times10^{-5}$. The $\pi(y)$ are left un-normalised 
in this figure for clarity.}
\label{fig:functions}
\end{figure}

In principle one can use a shooting method to find a 
trajectory which goes through any given point in the phase space, and hence 
the probability at that point.
In practice, because of the highly oscillatory nature of the solutions, it is 
easier to calculate cross-sections of the distribution by calculating where a 
large set of trajectories intersect a given line in the $x$-$y$ plane. The set 
of trajectories are computed by using initial conditions which lie on a small 
circle surrounding the fixed point~\cite{Dykman1994}. In this way a discretised
estimate of $\pi(\xx)$ can be computed. Figure \ref{fig:functions}a shows the 
functions $S(\xx)$ and $K(\xx)$ evaluated where trajectories intersect the 
line $x=0.06$. The probability density along this cross section is then 
given by $P_{\nn}\equiv\pi(\xx)=K(\xx)\exp[-N S(\xx)]$, shown in 
Figure \ref{fig:functions}b. Simulation results are not shown in 
Figure \ref{fig:functions}, but are in excellent agreement with the analytic 
results. Although there is no reason to suppose that these functions have
simple analytic forms, one could ask if a simple analytic expression can be
fitted in the asymptotic regime $y \gg y^{*}$. We have observed that in this
regime good fits to the functions shown in Figure \ref{fig:functions}a are
$K(y)\sim a/(b+y)$ and $S(y)\sim(c y+d)$, for some constants $a, b, c$ and $d$.
There is, however, no analytic justification for these.

For comparison, Figure \ref{fig:functions}c shows 
$\pi$ with and without the pre-factor $K$. Clearly for smaller systems, this 
contribution from the next-to-leading order terms in the expansion cannot be 
ignored when finding an accurate estimate of $\pi$.
More generally, we wish to calculate the full marginal infective probability 
density, $P_m=\sum_{n} P_{n,m}$. This is achieved by simply calculating the 
probability density along a range of cross-sections and then summing over them.
The resulting marginal density is then normalised such that $\sum_m P_m=1$, as 
it ought to be.

\section{Outbreak distributions}
\label{sec:results}

In this section we present results which show how the outbreak distribution 
changes with population size, $N$. The mean incidence varies linearly according
to $m^*=N y^*$. As the mean is decreased, three dynamical regimes can be 
distinguished, classified by population size, which coincide roughly with type 
I, II and III dynamics originally suggested by Bartlett~\cite{bartlett2}. 
Type I systems are large; they are those in which fade-out does not occur on 
a practical timescale, thus the disease is endemic. For the parameters used 
in this paper the lower limit on this class is $N\sim8\times 10^5$. For type 
II, medium size systems, fade-out can occur but reintroduction is swift and 
the dynamics are still mainly endemic. Type III are small systems which cannot 
support endemic infection levels. In these systems outbreaks are triggered by 
immigration events, which then rapidly lead to fade-out. Three time series 
representative of these regimes are shown in Figure \ref{fig:time_series}. The 
marginal infective distribution, $P_m$, then describes the distribution of 
outbreak sizes. We will primarily be interested in how this changes with $N$, 
for type I and II systems.

\begin{figure}[ht]
\centering
\includegraphics[width=0.8\textwidth]{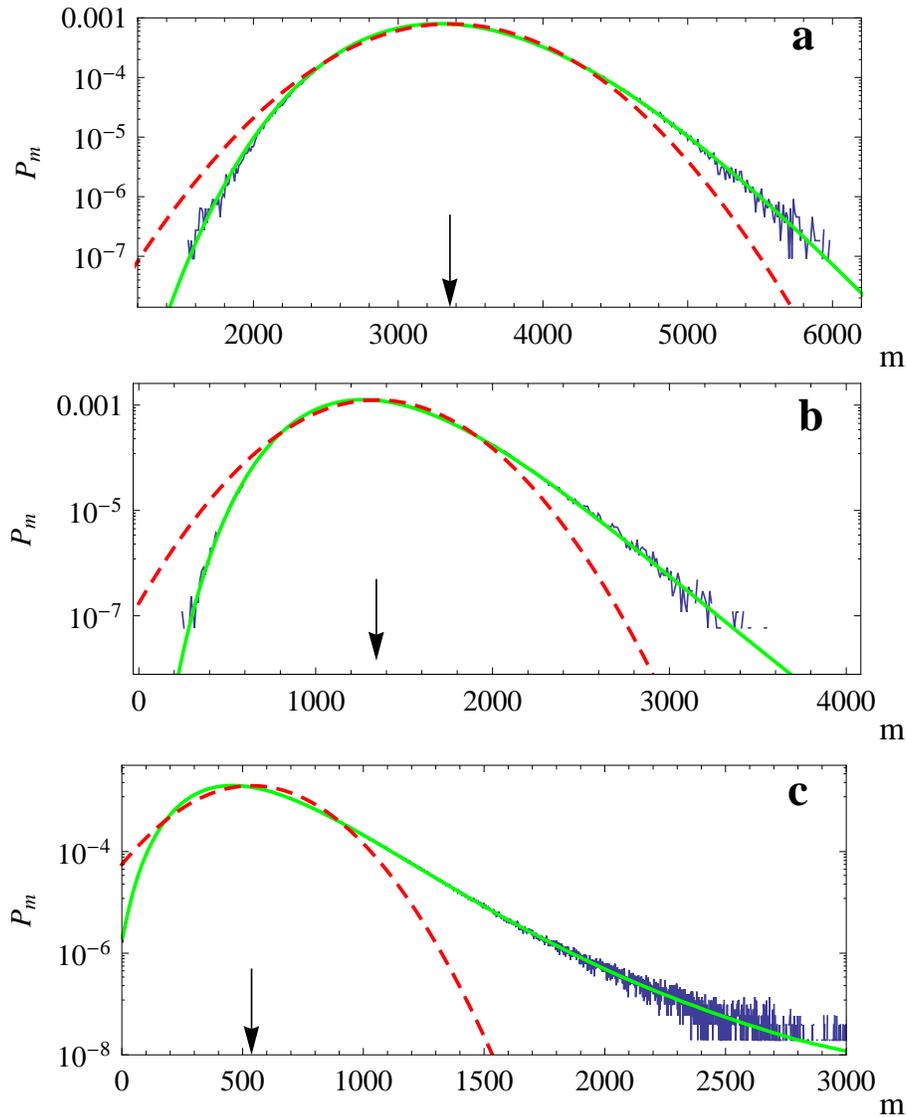}
\caption{Marginal infective probability densities for larger systems. 
Population sizes are (a) $N=5\times10^6$, (b) $N=2\times10^6$ and (c) 
$N=8\times10^5$. Solid green lines are from WKB calculations, the red dashed 
lines show the Gaussian approximation, and the noisy blue lines are from 
stochastic simulations. The arrows mark the mean values.} 
\label{fig:large}
\end{figure}

\subsection{Type I systems}
Figure \ref{fig:large} shows the outbreak distributions, calculated via the 
WKB approximation and compared to simulations, for three large populations in 
which the probability of extinction is negligible. For all three the agreement 
between the result from the WKB calculation and the stochastic simulations 
is excellent. As $N$ is decreased so is the mean (indicated by the arrows) and 
the distribution becomes more asymmetric as the fade-out boundary starts to 
exert a bigger influence on the dynamics. These distributions are also compared
with the Gaussian result given by \Eref{Gaussian_approx} (red dashed lines). 
It can be seen that this  overestimates the probability of the number of 
infected becoming small, while severely underestimating the probability of 
very large outbreaks. The WKB approximation remains valid down to 
$m\sim{\cal O}(1)$, but diverges close to the boundary as  
expected~\cite{Assaf2010}. 

\begin{figure}[ht]
\raggedleft
\includegraphics[width=0.9\textwidth]{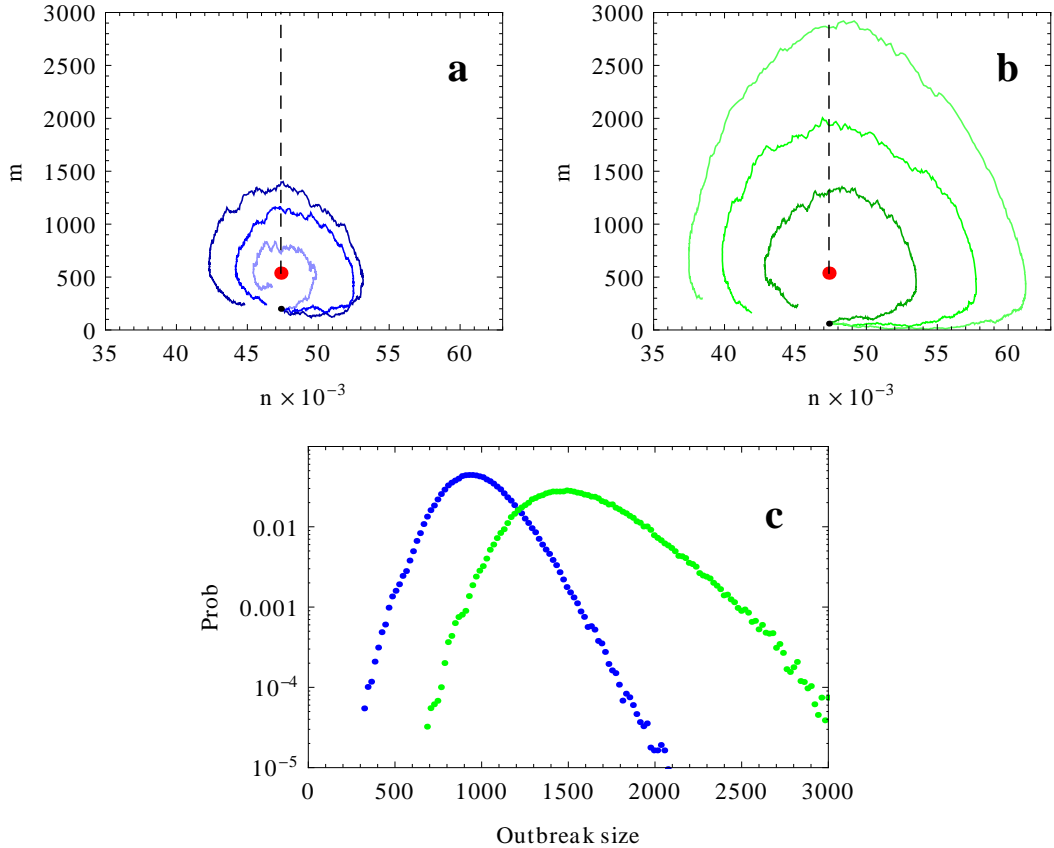}
\caption{The effect of the boundary on outbreak size. Parts (a) and (b) show 
realisations of the stochastic model started from $\xx_0=(Nx^*,200)$ and 
$\xx_0=(Nx^*,60)$ respectively, with $N=8\times10^5$. Dynamics then proceed in 
an anti-clockwise fashion. The realisations that fluctuate nearer to the 
boundary tend to move further along the $n$-axis leading to larger outbreaks. 
Part (c) shows the outbreak distributions---defined as when the realisation 
first crosses the dashed lines---for the two initial conditions.} 
\label{fig:outbreak}
\end{figure}

The fluctuations when $m$ is small (usually after an outbreak) determine the 
size of the next outbreak. If a fluctuation takes the system closer to the 
boundary then the probability that the next outbreak will be large is greatly 
increased. This is because if $m$ remains small, a large pool of susceptibles 
can build up. This effect is illustrated in Figure \ref{fig:outbreak}, which 
shows how the outbreak distribution changes with initial conditions close to 
and further away from the boundary. Figure \ref{fig:outbreak}b also shows that 
larger outbreaks have a naturally longer period. This feature accounts for the 
observed changes in the power spectrum of fluctuations in smaller systems, 
which are systematically shifted to lower frequencies, as compared with the 
$N\to\infty$ predictions~\cite{Simoes2008,Black2010}.

\subsection{Type II and III systems}
\label{sec:type2}
\begin{figure}[ht]
\centering
\includegraphics[width=0.8\textwidth]{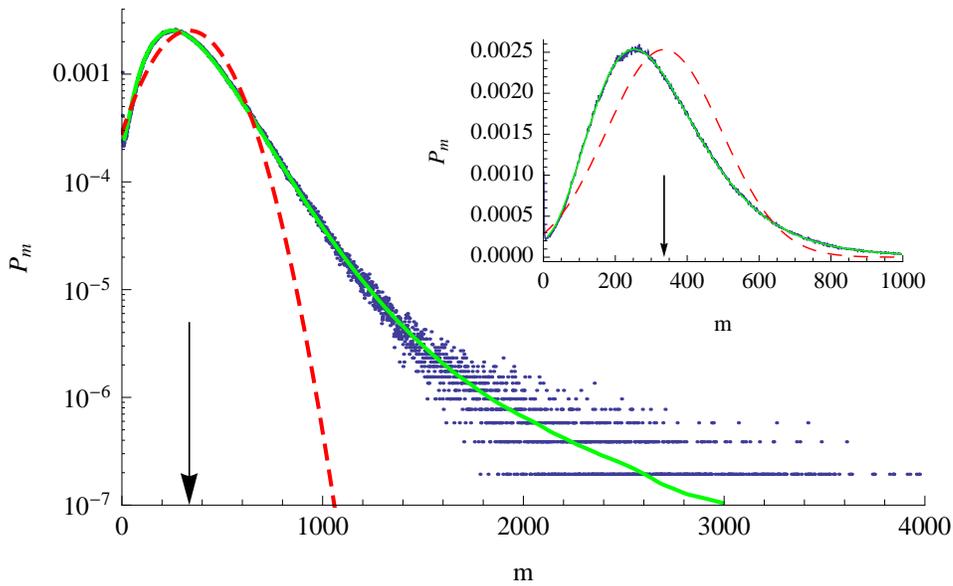}
\caption{Outbreak distribution for $N=5\times 10^5$. Solid green line is the 
WKB result, red dashed line is the Gaussian approximation and the blue dots 
are from simulations. Inset shows the asymptote around the fixed point; the 
probability of fade out is now clearly non-zero. The arrow marks the mean 
value.} 
\label{fig:medium}
\end{figure}

For the population sizes considered in the previous section, fade-out for any 
length of time is very rare, but even so, the proximity of the boundary and 
the potentially low levels of infectious individuals can have a large effect 
on the dynamics. We now study two smaller systems, and show how fade-outs of 
the disease affect the dynamics, and where the WKB approximation starts to 
break down. The outbreak distribution for $N=5\times 10^5$ is shown in Figure 
\ref{fig:medium}. The inset shows the asymptote of the distribution near the 
fixed point and there is clearly a non-zero probability of being on the 
boundary. In spite of this, the WKB result is still in very good agreement 
with the simulation result.

\begin{figure}[ht]
\centering
\includegraphics[width=0.8\textwidth]{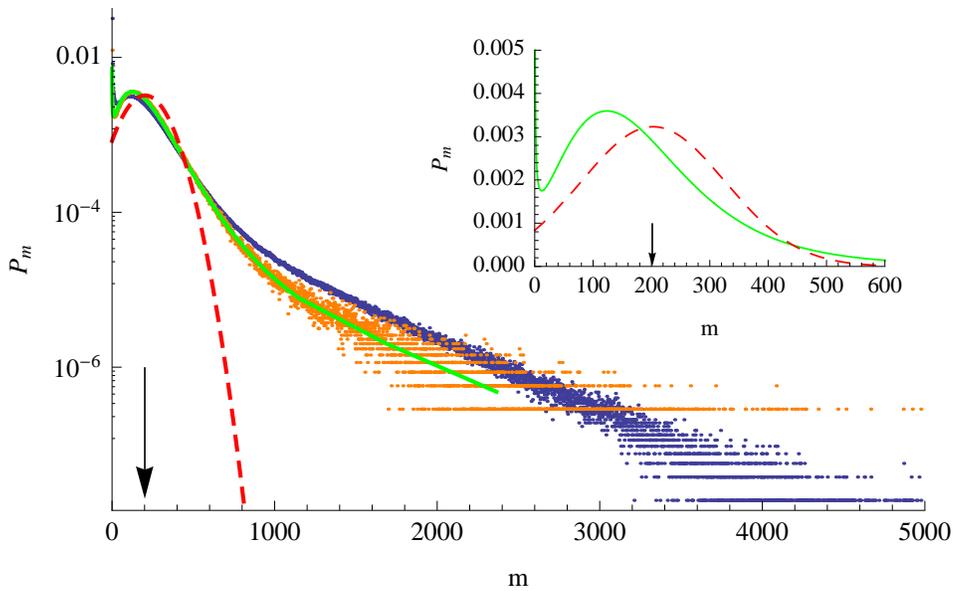}
\caption{Comparison of the full probability density (blue dots) and a single 
cross-section through the middle of the distribution (orange dots) for 
$N=3\times 10^5$. Again the green line, red dashed line and arrow show the WKB,
Gaussian and mean result respectively. At this system size the WKB 
approximation cannot capture the full distribution, but it does approximate 
well the central cross-section. The inset shows the asymptote of the 
cross-section about the fixed point, which shows that there is significant 
build up of probability on the boundary.} 
\label{fig:small}
\end{figure}

For smaller populations this agreement starts to break down. This can be seen 
in Figure \ref{fig:small} which shows the outbreak distribution for 
$N=3\times 10^5$. Although the WKB method provides a good approximation to a 
cross-section through the middle of the distribution, it cannot correctly 
predict the tails of the full marginal distribution, which is now strongly 
bimodal. This is because a large contribution to these tails comes from 
outbreaks which are started {\em on} the boundary. Thus to estimate the tails 
by the WKB method one would need to find trajectories which start on the 
boundary. This might in principle be possible---if we could generate initial 
conditions. 

For the larger system in Figure \ref{fig:medium} there are also outbreaks which 
originate from the boundary but their contribution to the overall marginal 
density can still be calculated from trajectories started at the fixed point. 
This is because the system never moves too far along the boundary before an 
immigration event starts a new outbreak. For $N=3\times10^5$ this is no longer 
the case as the system can potentially move far along the boundary into a 
region where it is no longer possible to find trajectories via the normal 
methods. 

The dynamics of even smaller systems are dominated by fade-out at the boundary 
as shown in Figure \ref{fig:time_series}c. In effect, almost all outbreaks 
are started from the boundary and are of such a large size that they almost 
always lead to fade-out. Susceptibles then build up through births until an 
immigration event triggers another outbreak. Therefore the timing of an 
outbreak is a random process, with the average outbreak size related to the 
time since the last outbreak (and hence the size of the pool of susceptibles).


\section{Discussion}
\label{sec:dis}
In this article we have analytically investigated the outbreak distribution 
of a stochastic SIR model with immigration. This is carried out by calculating 
both the leading order exponential, and next-to-leading order pre-factor in a 
WKB expansion of the master equation. The agreement of these analytic results 
with stochastic simulations is excellent. Although we cannot calculate the 
full marginal distribution for very small systems, where fade-out starts to 
dominate the dynamics, the WKB calculation is still remarkably accurate when 
calculating cross-sections of the bulk of the distribution. 

The ability of smaller systems---those where the number of infected individuals
can become small---to generate relatively larger outbreaks is clear in 
simulations and is important for analysing real time series. Up until now this 
effect has received little theoretical attention. In the limit $N\to\infty$ the 
distribution is Gaussian, a fact which is the basis of the system-size 
expansion~\cite{van_kampen}. But as $N$ is made smaller, the distribution 
becomes highly non-Gaussian as the fade-out boundary approaches the macroscopic
fixed point. Although the asymptote of the distribution remains somewhat 
Gaussian in shape, the fat tails of the distribution indicate the presence of 
large outbreaks (relative to the mean infectious level). These tails cannot be 
captured within the Gaussian approximation. Clearly these results are 
relevant to a number of models of this type; the SIR model is just one 
example of a generalised predator-prey model \cite{Brauer2001}. 

For smaller systems, the next-to-leading order term makes a significant 
contribution to the calculation of the probability density (see Figure 
\ref{fig:functions}c). Most studies of two-dimensional systems up until now 
have obtained approximate results for fixation times in various limits where 
Hamilton's equations can be simplified and solved 
analytically~\cite{Kamenev2008,MS09,Khasin2010}. Although we have not been 
concerned with fixation in this paper, our work has implications for 
calculating the mean time to extinction using these methods. By definition, if 
a system is small enough that extinction can happen on an observable time 
scale, then there will be large corrections to the first-order result from 
the next-to-leading order terms in the WKB expansion. 

Throughout this paper we have fixed the basic parameters of the system 
corresponding to a disease like measles, which has highly oscillatory 
stochastic dynamics and is typical of many childhood diseases~\cite{and_may}. 
Our aim has been to illustrate the method which we use, giving enough 
detail so that others can use it for other parameter values or other models. 
The choice of parameters was therefore not chosen on theoretical grounds; in 
principle any set which showed similar effects could have been used. While 
this is true for the standard SIR parameters $\beta$, $\gamma$ and $\mu$, it is 
worthwhile stressing the role of the parameter $\eta$, which is less standard. 
This parameter governs the rate of new infections brought back by susceptibles 
visiting outside populations. In the natural dynamics this can be thought of 
as a damping term because it is constantly depleting the pool of susceptibles, 
thus the larger $\eta$ is set, the smaller and less coherent the size of the 
stochastic oscillations for a given population size~\cite{alonso}. Such a 
system will show the same overall changes in the form of the outbreak 
distribution, but the onset will be later at comparatively smaller system 
sizes. 

One of the factors we have ignored in this paper is the influence of seasonal 
forcing on the epidemic dynamics. This is known to be very important for 
childhood diseases and arises due to the aggregation of children in schools 
during term time. To include this, the transmission term $\beta$ is made a 
function of time with a period of one year. The forcing then induces a limit 
cycle in the macroscopic dynamics with stochastic amplification of the 
transients about this~\cite{Black2010}. One of the most interesting and 
biologically relevant features of these forced models is the existence of 
multiple stable attractors; for example the coexistence of 1 and 2-year 
cycles~\cite{earn_science,Sch85b}. The WKB methods outlined in this article 
should be suitable for analysing the rate of switching between these attractors
and for showing how this changes with system size~\cite{Smelyanskiy1997}. A 
quantitative stochastic theory of this phenomenon would be a valuable addition 
to our overall understanding of these epidemic systems. 

\ack
A.J.B. acknowledges support from the EPSRC and the Australian Research 
Council's Discovery Projects funding scheme (project number DP110102893).

\section*{References}

\bibliographystyle{iopart-num}
\bibliography{wkb_refs.bib}

\end{document}